\documentclass{PoS}

\title{
\vspace*{-1.9cm}
\begin{minipage}{\textwidth}
{\normalfont\small Nikhef 2018-003, LTH 1149
\hspace{\fill} January 2018}\\
\end{minipage}\\[90pt]
 The R$^*$-operation and five-loop calculations}

  \ShortTitle{$R^*$ and five-loop calculations}
  
  \author{\speaker{Ben Ruijl}\\
  Institute for Theoretical Physics, ETH Zurich, Wolfgang-Pauli-Str. 27, 8093 Z\"urich, CH\\
  E-mail: \email{bruijl@ethz.ch}}

  \author{Franz Herzog, Takahiro Ueda, Jos Vermaseren\\
  Nikhef Theory Group, Science Park 105, 1098 XG Amsterdam, NL\\
  E-mails: \email{fherzog@nikhef.nl, tueda@nikhef.nl, t68@nikhef.nl}}

  \author{Andreas Vogt\\
  Department of Mathematical Sciences, University of Liverpool, Liverpool L69 3BX, UK\\
  E-mail: \email{Andreas.Vogt@liv.ac.uk}
  \\ \\ \\}
  
\usepackage{bm}
\usepackage{epsfig}
\usepackage{cite}
\usepackage{tikz}
\usepackage{amsmath}
\usepackage{mathtools}
\usepackage{url}
\usepackage{verbatim}
\usepackage{axodraw2}
\usetikzlibrary{calc,shapes.multipart,chains,arrows,fit,shapes,calc,backgrounds,trees,
decorations.pathreplacing,decorations.markings,intersections,patterns,positioning,matrix}

\newcommand{\ax}[5]{ \tikz[baseline=-2.5pt,inner sep=0pt,outer sep=0pt]{\node[] at (0,0) {\begin{axopicture}(#1,#2)(#3,#4) #5 \end{axopicture}};} }
\newcommand{\axb}[5]{ \tikz[baseline=+2.5pt,anchor=south,inner sep=0pt,outer sep=0pt]{\node[] at (0,0) {\begin{axopicture}(#1,#2)(#3,#4) #5 \end{axopicture}};} }

\newcommand{\beq}{\begin{equation}}
\newcommand{\eeq}{\end{equation}}
\newcommand{\bea}{\begin{eqnarray}}
\newcommand{\eea}{\end{eqnarray}}
\newcommand{\nin}{\noindent}
\newcommand{\nn}{\nonumber}
\newcommand{\MSb}{$\overline{\mbox{MS}}$}
\newcommand{\ra}{\rightarrow}
\newcommand{\ep}{\varepsilon}

\newcommand{\als}{\alpha_{\text s}}
\newcommand{\ars}{a_{\text s}}

\def\z#1{{\zeta_{#1}^{}}}

\def\MHt{{M_{H}^{\,3}}}
\def\MHs{{M_{H}^{\,2}}}
\def\MH{{M_{H}}}
\def\MZ{{M_{Z}}}

\def\qs{{q^{\:\!2}}}

\def\beq{\begin{equation}}   
\def\eeq{\end{equation}}
\def\bea{\begin{eqnarray}}  
\def\eea{\end{eqnarray}} 
\def\nn{\nonumber}
\def\r{\right} 
\def\l{\left} 
\def\f21{{}_2F_{1}}

\def\nl{{n^{}_{\! f}}}

\def\UVlineCol{Blue}

\newcommand{\MM}{\text{MM}}

\newcommand{\FORM}{\textsc{Form}}

\def\as(#1){{\alpha_{\text s}^{\,#1}}}
\def\ar(#1){{a_{\text s}^{\,#1}}}

\def\z#1{{\zeta_{\:\!#1}^{}}}

\def\nf{{n^{}_{\! f}}}

\newcommand{\equal}{\:\: = \:\:}

\def\bubbleonenum#1#2{
\raisebox{-7pt}
{
\begin{axopicture}{(30,20)(-13,-10)}
\SetScale{1}\SetColor{\UVlineCol}%
\Line(-15,0)(-10,0)
\Line(10,0)(15,0) 
\CArc(0,0)(10,0,360)
\Vertex(-10,0){1.5}
\Vertex(10,0){1.5}
\SetScale{1}\SetColor{Black}%
\Text(1.5,6.5){\tiny $#1$ \tiny}
\Text(1.5,-6.5){\tiny $#2$ \tiny}
\end{axopicture}
}
}

\def\bubbletwonum#1#2#3#4{
\raisebox{-7pt}
{
\begin{axopicture}{(30,20)(-13,-10)}
\SetScale{1}\SetColor{\UVlineCol}%
\Line(-15,0)(-10,0)
\Line(10,0)(15,0) 
\CArc(0,0)(10,0,360)
\CArc(10,10)(10,180,270)
\Vertex(-10,0){1.5}
\Vertex(10,0){1.5}
\Vertex(0,10){1.5}
\SetScale{1}\SetColor{Black}%
\Text(-5,4){\tiny $#1$ \tiny}
\Text(2,1.5){\tiny $#2$ \tiny}
\Text(6,5.5){\tiny $#3$ \tiny}
\Text(0,-7){\tiny $#4$ \tiny}
\end{axopicture}
}
}

\def\sunrisedot#1#2#3{
\raisebox{-7pt}
{
\begin{axopicture}{(25,25)(-11,-10)}
\SetScale{1}\SetColor{\UVlineCol}%
\Line(-15,0)(-10,0)
\Line(10,0)(15,0)
\Line(-10,0)(10,0) 
\CArc(0,0)(10,0,360)
\Vertex(-10,0){1.5}
\Vertex(10,0){1.5}
\Vertex(0,-10){1.5}
\SetScale{1}\SetColor{Black}%
\Text(1.5,6.5){\tiny $#1$ \tiny}
\Text(1.5,-4){\tiny $#2$ \tiny}
\Text(1.5,-14){\tiny $#3$ \tiny}
\end{axopicture}
}
}

\def\bubblethreenum#1#2#3#4#5#6{
\raisebox{-7pt}
{
\begin{axopicture}{(30,20)(-13,-10)}
\SetScale{1}\SetColor{\UVlineCol}%
\Line(-15,0)(-10,0)
\Line(10,0)(15,0) 
\CArc(0,0)(10,0,360)
\CArc(10,10)(10,180,270)
\CArc(-10,-10)(10,0,90)
\Vertex(-10,0){1.5}
\Vertex(0,-10){1.5}
\Vertex(10,0){1.5}
\Vertex(0,10){1.5}
\SetScale{1}\SetColor{Black}%
\Text(-9,8){\tiny $#1$ \tiny}
\Text(6,5){\tiny $#2$ \tiny}
\Text(10,9){\tiny $#3$ \tiny}
\Text(10,-9){\tiny $#4$ \tiny}
\Text(-9,-9){\tiny $#5$ \tiny}
\Text(-4,-5){\tiny $#6$ \tiny}
\end{axopicture}
}}

\def\vacsunrisedot#1#2#3{
\raisebox{-7pt}
{
\begin{axopicture}{(25,20)(-11,-10)}
\SetScale{1}\SetColor{\UVlineCol}%
\Line(-10,0)(10,0) 
\CArc(0,0)(10,0,360)
\Vertex(-10,0){1.5}
\Vertex(10,0){1.5}
\Vertex(0,-10){1.5}
\SetScale{1}\SetColor{Black}%
\Text(1.5,6.5){\tiny $#1$ \tiny}
\Text(1.5,-4){\tiny $#2$ \tiny}
\Text(1.5,-14){\tiny $#3$ \tiny}
\end{axopicture}
}
}

\def\vacsunrisedotdot#1#2#3{
\raisebox{-7pt}
{
\begin{axopicture}{(25,20)(-11,-10)}
\SetScale{1}\SetColor{\UVlineCol}%
\Line(-10,0)(10,0) 
\CArc(0,0)(10,0,360)
\Vertex(-10,0){1.5}
\Vertex(10,0){1.5}
\Vertex(-4,-9){1.5}
\Vertex(4,-9){1.5}
\SetScale{1}\SetColor{Black}%
\Text(1.5,6.5){\tiny $#1$ \tiny}
\Text(1.5,-4){\tiny $#2$ \tiny}
\Text(1.5,-14){\tiny $#3$ \tiny}
\end{axopicture}
}
}

\def\no1example{
\raisebox{-26pt}
{
\begin{axopicture}{(150,60)(10,0)}
\SetScale{0.75}
\Line(150,10)(70,70)
\SetWidth{4} \SetColor{White}
\Line(110,70)(70,10)
\Line(150,70)(110,10)
\SetWidth{0.5} \SetColor{Black}
\Line(110,70)(70,10)
\Line(150,70)(110,10)
\Line(40,40)(20,40)
\Arc(70,40)(30,90,180)
\Arc(70,40)(30,180,270)
\Line(110,70)(70,70)
\Line(70,10)(110,10)
\Line(150,70)(110,70)
\Line(110,10)(150,10)
\Arc(150,40)(30,0,90)
\Arc(150,40)(30,270,360)
\Line(200,40)(180,40)
\Vertex(40,40){1.5}
\Vertex(70,70){1.5}
\Vertex(70,10){1.5}
\Vertex(110,70){1.5}
\Vertex(110,10){1.5}
\Vertex(150,70){1.5}
\Vertex(150,10){1.5}
\Vertex(180,40){1.5}
\Vertex(82,70){1.5}
\Vertex(94,70){1.5}
\Vertex(174,58){1.5}
\Vertex(177,27){1.5}
\Vertex(166,15){1.5}
\Text(47,15)[rt]{\small $\rho \sigma$ \small}
\Text(176,18)[lt]{\small $\kappa \lambda$ \small}
\Text(128,72)[b]{\small $\mu \nu$ \small}
\Text(93,5)[t]{\small $\mu \nu$ \small}
\Text(125,5)[t]{\small $\rho \sigma$ \small}
\Text(138,40)[l]{\small $\kappa \lambda$ \small}
\end{axopicture}
}
}

\abstract{ 
We sketch how the $R^*$-operation can be used to compute the pole terms of
Feynman diagrams. We identify computational difficulties when performing 
five-loop calculations, and provide four solutions that drastically reduce the
number of terms that are generated. Using these methods, we have computed the 
beta function for Yang-Mills theory with fermions, the $R$-ratio in 
electron-positron annihilation, and Higgs decays to quarks and gluons at 
five-loop accuracy. The results for the beta function and Higgs decay width
to gluons in the heavy-top limit are briefly discussed.
There is no need for six-loop extensions of these calculations in the 
near future.
}

\FullConference{13th International Symposium on Radiative Corrections\\
24-29 September 2017, St.\ Gilgen, Austria}

\begin{document}

\section{Introduction}
\vspace*{-1mm}

\nin
Many physically interesting quantities can be computed from the pole parts of Feynman diagrams,
for example, anomalous dimensions (such as the beta function), splitting functions, and decay rates.
Often, it is much easier to compute the pole parts of diagrams than the finite part.

\vspace*{0.8mm}
In this work we sketch the $R^*$-operation~\cite{Chetyrkin:1982nn,Chetyrkin:1984xa,Herzog:2017bjx}, 
which is able to compute the pole parts of $L$-loop diagrams by a computation of at most 
$(L\!-\!1)$-loop diagrams. 
Using a high-performance implementation of the $R^*$-operation combined with the \textsc{Forcer} 
program~\cite{tuLL2016,ForcerCode,FORCER}, we are able to compute the poles of five-loop massless 
propagator diagrams. Our programs rely heavily on \FORM{}~\cite{FORM3,TFORM,FORM4,Ruijl:2017dtg}.

\vspace*{0.8mm}
Performing efficient computations of five-loop diagrams is hard, since the Feynman rules create 
many terms.
To alleviate these issues, we describe four methods to reduce the number of~terms which are not 
specific to the $R^*$-method: 
(1) removal of propagator insertions, (2) delaying of Feynman rule substitution, 
(3) canonicalization of Feynman diagrams, and (4) efficient tensor reduction. 

\vspace*{0.8mm}
Using these optimizations, the five-loop beta function for Yang-Mills theory with fermions 
\cite{Herzog:2017ohr} can be computed in six days on one 32-core machine. We have also computed 
the $R$-ratio in $e^+e^-\rightarrow$ hadrons, and the Higgs decays to bottom quarks and 
gluons at five loops~\cite{Herzog:2017dtz}.

\vspace*{0.8mm}
The outline of this paper is as follows. 
In section~\ref{sec:rstar} we very briefly describe the $R^*$-operation.
In section \ref{sec:optimisations} we address the four optimizations.
In section~\ref{sec:results} we briefly discuss the above physics results.
Finally, we summarize and present a brief outlook in~\ref{sec:conclusion}.

\vspace*{-1mm}
\section{The $R^*$-operation}
\vspace*{-1mm}
\label{sec:rstar}

\nin
The $R^*$-operation can be used to compute the poles of Feynman 
diagrams~\cite{Chetyrkin:1982nn,Chetyrkin:1984xa}. Recently, it has been
extended to Feynman diagrams with arbitrary numerator structure~\cite{Herzog:2017bjx}.
In this section we briefly sketch how the $R^*$-operation works, focusing on UV-counterterms.

\vspace*{0.5mm}
The basic object of the $R^*$-operation is the UV counterterm operation $\Delta$ acting on a graph 
$G$, which is defined as the poles of $G$ in the limit of all loop momenta going to infinity with 
all contributions from subdivergences subtracted.
Additionally, we define the pole operator $K$ for a Laurent series in $\ep$
\begin{equation}
K \sum_{i=-\infty}^\infty c_i \, \ep^i = \sum_{i=-\infty}^{-1} c_i \, \ep^i \;.
\end{equation}
Then the $R^*$ operation for some simple examples yields
\begin{align}
K \bubbleonenum{1}{2}&= \Delta \l(\bubbleonenum{1}{2}\r)\\
K \bubbletwonum{1}{2}{3}{4} &= \Delta \l(\bubbletwonum{1}{2}{3}{4}\r) 
 + \Delta \l(\bubbleonenum{2}{3}\r) \cdot \bubbleonenum{1}{4} \;.
\end{align}
In general, all sets of non-overlapping divergent subdiagrams have to be considered:
\begin{align}
K\bubblethreenum{1}{2}{3}{4}{5}{6} &= \Delta \l(\bubblethreenum{1}{2}{3}{4}{5}{6}\r) \nn\\
& +\Delta \l(\bubbleonenum{5}{6}\r) \cdot \bubbletwonum{1}{2}{3}{4} 
  +\Delta \l(\bubbleonenum{2}{3}\r) \cdot \bubbletwonum{4}{5}{6}{1} \\
& - \Delta \l(\bubbleonenum{5}{6}\r) \Delta \l(\bubbleonenum{2}{3}\r) 
  \cdot \bubbleonenum{1}{4}\nn \;.
\end{align}
Here the contribution from two counterterms gets a minus sign, to prevent double counting.

\pagebreak

For logarithmically divergent diagrams, $\Delta$ does not depend on external momenta or masses. 
Consequently, we can infrared rearrange (IRR)~\cite{Vladimirov:1979zm}:
\begin{equation}
\Delta \l(\bubbletwonum{}{}{}{}\r) = \Delta \l(\vacsunrisedot{}{}{}\r) 
  = \Delta \l(\sunrisedot{}{}{}\r) \;.
\end{equation}
Using IRR and the definition of $\Delta$, we can express the UV-counterterm of $G$ in terms 
of simpler diagrams:
\begin{equation}
\Delta(G) \mathrel{\overset{\makebox[0pt]{\mbox{\normalfont\tiny\sffamily IRR}}}{=}} 
  \underbrace{\Delta (G')}_{\text{Simpler than G}} 
  \!\! = \;  K(G') - \underbrace{\text{subdivergences}(G')}_{\text{Lower-loop diagrams}}
\end{equation}
 
Using this setup, we can rewrite all $L$-loop diagrams to $(L\!-1\!)$-loop scalar massless 
propagator integrals. For five-loop applications, all those integrals can be computed using
{\sc Forcer}~\cite{ForcerCode,FORCER}.

\vspace*{-1mm}
\section{Optimisations}
\label{sec:optimisations}
\vspace*{-1mm}

\nin
Performing computations at five loops introduces at least four new bottlenecks compared to four 
loops:
(1) the number of diagrams and their complexity grow exponentially,
(2) the substitution of the Feynman rules is slow and creates millions of terms,
(3) the number of counterterms grows exponentially, and
(4) tensors of rank 10 have to be reduced, which involves solving large systems.

\vspace*{0.5mm}
In this section we address these issues by presenting five optimisations,
namely improved treatment of propagator insertions in section~\ref{sec:selfenergy},
delayed Feynman rule substitution in section~\ref{sec:delayed},
a canonical form algorithm for Feynman diagrams in section~\ref{sec:canonical},
and an efficient tensor reduction algorithm in section~\ref{sec:tensor}.

\subsection{Treatment of propagator insertions}
\label{sec:selfenergy}
\vspace*{-1mm}

\nin
Many of the higher-loop corrections are self-energies of propagators in the diagram.
Due to the local nature of the Feynman rules, these self-energies only depend on their external 
momentum (there are no contractions
with other parts of the larger diagram), so they can be `factorised' out,
\begin{align}
\begin{split}
&\axb{80}{50}{10}{45}{
\SetArrowScale{0.75}
\PhotonArc(50,50)(25,0,60){2}{4}
\PhotonArc(50,50)(25,120,180){-2}{4}
\Photon(50,87.5)(50,62.5){-2}{4}
\CArc[arrow,flip](50,75)(12.5,0,90)
\CArc[arrow,flip](50,75)(12.5,90,180)
\CArc[arrow,flip](50,75)(12.5,180,270)
\CArc[arrow,flip](50,75)(12.5,270,360)
\Line[arrow,flip](25,50)(75,50)
\Line[arrow](87.5,50)(75,50)
\Line[arrow](25,50)(12.5,50)
}
+
\axb{80}{50}{10}{45}{
\SetArrowScale{0.75}
\PhotonArc(50,50)(25,0,60){2}{4}
\PhotonArc(50,50)(25,120,180){-2}{4}
\PhotonArc(50,87.5)(6.25,195,345){-1}{4}
\CArc[arrow,flip](50,75)(12.5,0,60)
\CArc[arrow,flip](50,75)(12.5,60,120)
\CArc[arrow,flip](50,75)(12.5,120,180)
\CArc[arrow,flip](50,75)(12.5,180,360)
\Line[arrow,flip](25,50)(75,50)
\Line[arrow](87.5,50)(75,50)
\Line[arrow](25,50)(12.5,50)
}
+
\axb{80}{50}{10}{45}{
\SetArrowScale{0.75}
\PhotonArc(50,50)(25,0,60){2}{4}
\PhotonArc(50,50)(25,120,180){-2}{4}
\PhotonArc(50,50)(25,70,105){-2}{2}
\CCirc(64.33,70.478){9}{White}{White}
\CCirc(35.7,70.478){9}{White}{White}
\CArc[arrow,flip](64.33,70.478)(9,315,495)
\CArc[arrow,flip](64.33,70.478)(9,495,675)
\CArc[arrow,flip](35.7,70.478)(9,45,225)
\CArc[arrow,flip](35.7,70.478)(9,225,405)
\Line[arrow,flip](25,50)(75,50)
\Line[arrow](87.5,50)(75,50)
\Line[arrow](25,50)(12.5,50)
\Vertex(25,50){1}
\Vertex(75,50){1}
}
=\\[-1mm]
&\left(
\ax{55}{35}{22}{57}{
\SetArrowScale{0.75}
\Photon(27.5,75)(37.5,75){-2}{2}
\Photon(62.5,75)(72.5,75){-2}{2}
\Photon(50,87.5)(50,62.5){-2}{4}
\CArc[arrow,flip](50,75)(12.5,0,90)
\CArc[arrow,flip](50,75)(12.5,90,180)
\CArc[arrow,flip](50,75)(12.5,180,270)
\CArc[arrow,flip](50,75)(12.5,270,360)
}
+
\ax{55}{35}{22}{57}{
\SetArrowScale{0.75}
\Photon(27.5,75)(37.5,75){-2}{2}
\Photon(62.5,75)(72.5,75){-2}{2}
\PhotonArc(50,87.5)(6.25,195,345){-1}{4}
\CArc[arrow,flip](50,75)(12.5,0,60)
\CArc[arrow,flip](50,75)(12.5,60,120)
\CArc[arrow,flip](50,75)(12.5,120,180)
\CArc[arrow,flip](50,75)(12.5,180,360)
}
+
\ax{70}{35}{15}{52}{
\SetArrowScale{0.75}
\Photon(15,70.478)(85,70.478){-2}{8}
\CCirc(64.33,70.478){9}{White}{White}
\CCirc(35.7,70.478){9}{White}{White}
\CArc[arrow,flip](64.33,70.478)(9,315,495)
\CArc[arrow,flip](64.33,70.478)(9,495,675)
\CArc[arrow,flip](35.7,70.478)(9,45,225)
\CArc[arrow,flip](35.7,70.478)(9,225,405)
}
\right)
\times
\axb{80}{40}{10}{45}{
\SetArrowScale{0.75}
\PhotonArc(50,50)(25,0,180){2}{8}
\Line[arrow,flip](25,50)(75,50)
\Line[arrow](87.5,50)(75,50)
\Line[arrow](25,50)(12.5,50)
\Line[width=1.25](34,78)(39,72)
\Line[width=1.25](34,72)(39,78)
\Line[width=1.25](56,78)(61,72)
\Line[width=1.25](56,72)(61,78)
} =\\[-1mm]
&\Sigma_2^{\text{1PR}}
\axb{80}{40}{10}{45}{
\SetArrowScale{0.75}
\PhotonArc(50,50)(25,0,180){2}{8}
\Line[arrow,flip](25,50)(75,50)
\Line[arrow](87.5,50)(75,50)
\Line[arrow](25,50)(12.5,50)
\Line[width=1.25](34,78)(39,72)
\Line[width=1.25](34,72)(39,78)
\Line[width=1.25](56,78)(61,72)
\Line[width=1.25](56,72)(61,78)
} \; ,
\end{split}
\end{align}
where the $L$-loop self-energy is replaced by $({p^2})^{-\ep L}$ in the larger diagram
(marked by $L$ crosses). In a sense, the subdiagram is integrated out. The resulting simpler 
topology is multiplied by the one-particle-reducible $L$-loop self-energy.
Since the $L$-loop subdiagram is of lower order, these quantities have already been computed and
can easily be tabulated to prevent recomputations. For example, a five-loop
diagram may contain the expensive 4-loop gluon propagator as a subdiagram.

\vspace*{0.5mm}
For the $R^*$-operation, this representation has an issue: the non-integer power hides 
UV-diver\-gent subdiagrams, which should be subtracted. However, since the exact contents
of the $(p^2)^{-\ep L}$ is factorised out, we may replace it with \emph{any} $L$-loop subdiagram.
Therefore we choose the simplest configuration: $L$ scalar one-loop bubbles side by side.
Thus, for the $R^*$-operation we can remove propagator insertions by using the following relation:
\begin{equation}
\ax{80}{80}{10}{10}{
\CCirc(50,50){25}{Black}{White}
\CCirc(50,75){12.5}{Black}{White}
\Line(87.5,50)(75,50)
\Line(25,50)(12.5,50)
\Vertex(25,50){1.5}
\Vertex(75,50){1.5}
\Text (50,75) {$L$}
}
=
\frac{
\ax{60}{40}{20}{30}{
\Line(75,50)(62.5,50)
\Line(37.5,50)(25,50)
\Arc(50,50)(12.5,180,360)
\Arc(50,50)(12.5,0,180)
\Text (50,50) {$L$}
}
}
{
\left(
\ax{60}{40}{20}{30}{
\Line(75,50)(62.5,50)
\Line(37.5,50)(25,50)
\Arc(50,50)(12.5,180,360)
\Arc(50,50)(12.5,0,180)
}
\right)^L
}
\times
\ax{80}{70}{10}{15}{
\CArc(50,50)(25,135,180)
\CArc(50,50)(25,0,45)
\DashArc(50,50)(25,0,135){2}
\CCirc(64.33,70.478){5}{Black}{White}
\CCirc(35.7,70.478){5}{Black}{White}
\CCirc(50,75){4}{Black}{White}
\Arc(50,50)(25,180,360)
\Line(87.5,50)(75,50)
\Line(25,50)(12.5,50)
\Vertex(25,50){1.5}
\Vertex(75,50){1.5}
\Text(35.7,70.478) {\tiny 1 \tiny}
\Text(64.33,70.478) {\tiny $L$ \tiny}
} \; .
\end{equation}

\subsection{Delayed Feynman rule substitution}
\label{sec:delayed}
\vspace*{-1mm}

\nin
Substituting the Feynman rules creates many terms. For example, the following fully gluonic 
five-loop graph creates $12\,029\,521$ scalar integrals in the Feynman gauge:
\begin{equation}
\ax{160}{60}{0}{0}{
\SetScale{0.75}
\Gluon(40,40)(10,40){2}{6}
\GluonArc(70,40)(30,90,180){2}{8}
\GluonArc(70,40)(30,180,270){2}{8}
\Gluon(96.67,70)(70,70){-2}{4}
\Gluon(70,10)(96.67,10){-2}{4}
\Gluon(123.34,70)(96.67,70){-2}{4}
\Gluon(96.67,10)(123.34,10){-2}{4}
\Gluon(150,70)(123.34,70){-2}{4}
\Gluon(123.34,10)(150,10){-2}{4}
\GluonArc(150,40)(30,0,90){2}{8}
\GluonArc(150,40)(30,270,360){2}{8}
\Gluon(210,40)(180,40){2}{6}
\Gluon(70,10)(70,70){2}{8}
\Gluon(96.67,10)(96.67,70){2}{8}
\Gluon(123.34,10)(123.34,70){2}{8}
\Gluon(150,10)(150,70){2}{8}
\Vertex(40,40){1.5}
\Vertex(70,70){1.5}
\Vertex(70,10){1.5}
\Vertex(96.67,70){1.5}
\Vertex(96.67,10){1.5}
\Vertex(123.34,70){1.5}
\Vertex(123.34,10){1.5}
\Vertex(150,70){1.5}
\Vertex(150,10){1.5}
\Vertex(180,40){1.5}
}\; .
\label{eq:fiveloopgluon}
\end{equation}
\nin
The source of the blow-up is the Feynman rule for the triple gluon vertex, which can be 
written in the following way:
\begin{equation}
v_{3g}^{}(p_1^{\,\mu,a},p_2^{\,\nu,b},p_3^{\,\rho,c}) 
 = -i f^{\,abc} \left[ (p_1-p_2)^\rho g_{\mu \nu} + (2 p_2+p_1)^\mu g_{\nu \rho}
    x + (-2 p_1-p_2)^\nu g_{\mu \rho} \right] \, .
\end{equation}
Thus, for every vertex, six terms are created, of which some will evaluate to the same expression 
due to symmetries.
For all these terms, expensive operations such as Taylor expansions and divergent subgraph 
recognitions have to be performed.
However, 
these operations only depend on the momentum powers and are invariant under the way
the momenta contract. So, we rewrite the triple gluon vertex in a way that exposes the momenta, 
but keeps all the contractions unsubstituted:
\begin{equation}
v_{3g}^{}(p_1^{\,\mu,a},p_2^{\,\nu,b},p_3^{\,\rho,c})
    = - i f^{\,abc} p_1^{\,\sigma} t_3(\sigma,\nu,\rho,\mu) 
      + i f^{\,abc} p_2^{\,\sigma} t_3(\sigma,\mu,\rho,\nu) \;,
\end{equation}
where
\begin{equation}
t_3(\mu,\nu,\rho,\sigma) 
= g_{\mu \rho} g_{\nu \sigma}+g_{\mu \sigma} g_{\nu \rho}-2 g_{\mu \nu} g_{\rho \sigma} \;.
\end{equation}
After rewriting $v_{3g}^{}$ in terms of $t_3$, there are only $2^{10}=1024$ terms for the Feynman 
diagram in eq.~\eqref{eq:fiveloopgluon}.
We can keep our input in this compactified notation for as long as the actual contractions are not 
important, which is right until the tensor reduction.

\vspace*{0.5mm}
We define the operation $\circ$ that applies the remaining Feynman rules to all components of the 
$R^*$-operation, for example, 
\begin{align}
   t_3(\mu,\nu,\rho,\sigma) \circ \Delta \left( \vacsunrisedotdot{}{}{\mu \nu} \right)a
   \bubbleonenum{\rho}{\sigma} &=
   2\Delta \left(\vacsunrisedotdot{}{}{\mu \nu} \right) \bubbleonenum{\mu}{\nu}
   -2\Delta \left(\vacsunrisedot{}{}{} \right) \bubbleonenum{\rho}{\rho} \;.
\end{align}
We stress that for this particular case contraction is \emph{necessary}.

\vspace*{0.5mm}
Similar rules can be devised for the other vertices and for the trace of gamma matrices. 
At five loops, the substitution of $t_3$ and similar structures is an expensive part
of the calculation, since the number of generated terms is high.

\subsection{Canonical forms for Feynman diagrams}
\label{sec:canonical}
\vspace*{-1mm}

\nin
The $R^*$-operation applied to five-loop diagrams will create many counterterms. In order to
reduce computation time, it is important to compute the counterterms of a specific graph only
once. In~turn, this requires an efficient way to detect if two graphs are equal.
One straightforward option is to keep a list of all the graphs that have already been processed 
and test for
isomorphisms on every element of the list until one is found. If no match is found, the current
graph can be added to the list. The two downsides of this method are that (1) an isomorphism test
can be rather slow at five loops and (2) that the list of topologies grows rapidly.

\vspace*{0.5mm}
A better alternative is to construct a \emph{canonical form} of a graph. A canonical form
is an isomorphism of the graph that is designated as the smallest by some yet to be defined
measure. To test for equality, one can simply compare the canonical forms.
 Since isomorphy is first and foremost a property of the vertices, we give each
vertex a label from $1$ to $n$. For simplicity, let us consider a graph that has no
dot products and only has edges with power 1.

\vspace*{0.5mm}
We convert our graph to an edge representation:
\begin{align}
\ax{70}{45}{-35}{-22}{
\Line(-30,0)(-20,0)
\Line(20,0)(30,0) 
\CArc(0,0)(20,0,360)
\CArc(20,20)(20,180,270)
\Vertex(-30,0){1.5}
\Vertex(30,0){1.5}
\Vertex(-20,0){1.5}
\Vertex(20,0){1.5}
\Vertex(0,20){1.5}
\Text(-15,-4){\small 1}
\Text(15,-4){\small 3}
\Text(-4,12){\small 2}
\Text(-30,6){\small 0}
\Text(30,6){\small 4}
}
= e(0,1) e(1,2) e(2,3) e(2,3) e(1,3) e(3,4) \; .
\label{eq:edge}
\end{align}
Here, $e(n_1,n_2)$ is the edge function, in which we place the smallest vertex index
as the first argument. 
The \emph{edge list} is a lexicographically sorted list of edge functions, as is shown
in eq.~\eqref{eq:edge}.
Now we define the smallest isomorphism of a graph as the vertex labeling
for which the edge list is lexicographically smallest.\footnote{In our program, we use the 
internal (deterministic) sorting order 
of \FORM{} \cite{FORM3,TFORM,FORM4} to determine the smallest isomorphism instead.
The latest version of \FORM{} \cite{Ruijl:2017dtg} is required for our R$^*$ program.}

\vspace*{0.5mm}
We can easily extend the graph notation to a graph where propagators can have different powers, 
by introducing a third argument to the edge function $e$:
\begin{align}
\ax{70}{45}{-35}{-22}{
\Line(-30,0)(-20,0)
\Line(20,0)(30,0) 
\CArc(0,0)(20,0,360)
\CArc(20,20)(20,180,270)
\Vertex(-30,0){1.5}
\Vertex(30,0){1.5}
\Vertex(-20,0){1.5}
\Vertex(20,0){1.5}
\Vertex(0,20){1.5}
\Vertex(-14,14){1.5}
\Vertex(14,14){1.5}
\Text(-15,-4){\small 1}
\Text(15,-4){\small 3}
\Text(-4,12){\small 2}
\Text(-30,6){\small 0}
\Text(30,6){\small 4}
}
= e(0,1,1) e(1,2,2) e(2,3,1) e(2,3,2) e(1,3,1) e(3,4,1) \; ,
\end{align}
where we again make sure that the first two arguments of $e(n_1,n_2,\ldots)$ are sorted. 
To add support for dot products and tensors, we extend the edge function even further:
\begin{align}
\ax{70}{45}{-35}{-22}{
\Line(-30,0)(-20,0)
\Line(20,0)(30,0) 
\CArc(0,0)(20,0,360)
\CArc(20,20)(20,180,270)
\Vertex(-30,0){1.5}
\Vertex(30,0){1.5}
\Vertex(-20,0){1.5}
\Vertex(20,0){1.5}
\Vertex(0,20){1.5}
\Vertex(-14,14){1.5}
\Vertex(14,14){1.5}
\Text(-25,-6){\small $\mu$}
\Text(19,19){\small $\mu$}
\Text(-15,-4){\small $1$}
\Text(15,-4){\small $3$}
\Text(-4,12){\small $2$}
\Text(-30,6){\small $0$}
\Text(30,6){\small $4$}
}= e(0,1,1,\mu) e(1,2,2) e(2,3,1) e(2,3,2,\mu) e(1,3,1) e(3,4,1) \;.
\end{align}

\vspace*{0.5mm}
We define the canonical signs of the momenta such that they always flow from the smallest vertex 
label to the highest.
If a transformation changes the order, we flip the sign if the number of vectors in the momentum is odd:
\begin{align}
  e(2,1,n,\mu_1,\ldots,\mu_k) = (-1)^{k} e(1,2,n,\mu_1,\ldots,\mu_k) \, .
\end{align}
Finally, the momentum label $p_i$ of each edge is uniquely defined by the position $i$ of the edge 
in the edge list.

\vspace*{0.5mm}
Now that most properties of the Feynman integral are captured in the extended edge list and
we have defined which edge list is smallest, we use McKay's canonicalisation 
algorithm~\cite{McKay81} to efficiently rewrite the complete Feynman integral to canonical form. 
A simplified version of this algorithm is implemented in \FORM{} code.

\subsection{Efficient tensor reduction}
\label{sec:tensor}
\vspace*{-1mm}

\nin
It can be shown that the tensor reduction of ultraviolet and infrared 
subtraction terms, required for the $R^*$-operation, is equivalent
to the tensor reduction of tensor vacuum bubble integrals. In general 
tensor vacuum integrals can be reduced to linear combinations of products of 
metric tensors $g^{\mu\nu}$ whose coefficients are scalar vacuum integrals.
Specifically a rank $r$ tensor, $T^{\mu_1\dots\,\mu_r}$, is written as a 
linear combination of $n=r!/2^{(r/2)}/(r/2)!$ combinations 
of $r/2$ metric tensors with coefficients $c_\sigma$, i.e.,
\beq
T^{\mu_1\dots\,\mu_r}=\sum_{\sigma \in\, {}_2 S_{r}} c_\sigma 
 \,T^{\mu_1\dots\mu_r}_\sigma\,,\qquad
T^{\mu_1\dots\,\mu_r}_\sigma= g^{\mu_{\sigma(1)}\mu_{\sigma(2)}} 
\dots\, g^{\mu_{\sigma(r-1)}\mu_{\sigma(r)}}\,.
\eeq
Here we define ${}_2S_{r}$ as the group of permutations which do \emph{not} 
leave the tensor $T^{\mu_1\dots\,\mu_r}_\sigma$ invariant. The coefficients 
$c_\sigma$ can be obtained by acting onto the tensor $T^{\mu_1\dots\,\mu_r}$ 
with certain projectors $P_\sigma^{\mu_1\dots\mu_r}$, such that
\beq
c_\sigma=P_\sigma^{\:\!\mu_1\dots\,\mu_r} T_{\mu_1\dots\,\mu_r}\,.
\eeq
From this it follows that the orthogonality relation,
\beq
\label{eq:orthogonality}
 P_\sigma^{\:\!\mu_1\dots\,\mu_r} T_{\tau,\,\mu_1\dots\,\mu_r} 
 = \delta_{\sigma\tau}\,,
\eeq
must hold, where $\delta$ is the Kronecker-delta. Since the projector 
$P_\sigma^{\:\!\mu_1\dots\,\mu_r}$ of each tensor can also be written in terms 
of a linear combination of products of metric tensors, inverting an $n\times n$
matrix determines all the projectors. However, there are two issues.
The first is that the size of the matrix grows 
rather rapidly as $r$ increases.  Instead of solving an $n \times n$ linear 
system, the symmetry group of the metric tensors can be utilised to reduce the 
size of the system.
From eq.~(\ref{eq:orthogonality}) it follows that the projector $P_\sigma$ is in the 
same symmetry group (the group of 
permutations which leave it invariant) as $T_\sigma$.
For example, given a permutation $\sigma_1=(123...(r-1)r)$,
\beq
T_{\sigma_1}^{\mu_1\dots\,\mu_r} = g^{\mu_1\mu_2} g^{\mu_3\mu_4} 
 \dots\, g^{\mu_{r-1}\mu_r}\,.
\eeq
The corresponding projector $P_{\sigma_1}^{\mu_1\dots\,\mu_r}$ must be 
symmetric under interchanges of indices such as $\mu_1 \leftrightarrow 
\mu_2$, $(\mu_1,\mu_2) \leftrightarrow (\mu_3,\mu_4)$ and so on.
Grouping the metric tensors by the symmetry leads to the fact that 
$P_\sigma$ is actually written in a linear combination of a small number of 
$m$ tensors instead of $n$ ($m \le n$),
\beq
P_\sigma^{\mu_1\dots\,\mu_r} =\sum_{k=1}^m b_k 
 \sum_{\tau\in A_m^\sigma} T^{\mu_1\dots\,\mu_r}_\tau.
\eeq
The set of groups $\{A_k^\sigma|k=1..m\}$ must therefore each be closed 
under the permutations which leaves $T_\sigma$ invariant and at the same 
time their union must cover once the group ${}_2S_n$. Contracting $P_\sigma$ 
with a representative in each group gives an $m \times m$ matrix which can be 
inverted to yield the coefficients $b_k$. 
The number of unknowns $m$ is $m=5$ for $r=8$ and $m=22$ for 
$r=16$, whereas we have $n=105$ for $r=8$ and $n=2027025$ for $r=16$. 
The comparison of these numbers illustrates that the exploitation of the 
symmetry of the projectors makes it possible to find the tensor reduction 
even for very large values of~$r$, which could never have been obtained 
by solving the $n\times n$ matrix.

\vspace*{0.5mm}
The second issue with tensors of high rank is the large number of intermediate terms
that are created. Even though the system for the projector can be solved efficiently, 
$\mathcal{O}(n^2)$ terms
will be created, of which some will merge due to symmetry. Let us consider rank 6, with 15 terms:
\begin{equation}
  c_1 g^{\mu_1 \mu_2}g^{\mu_3 \mu_4}g^{\mu_5 \mu_6}
  + c_2 g^{\mu_1 \mu_3}g^{\mu_2 \mu_4}g^{\mu_5 \mu_6} + \ldots \; .
\label{eq:rank6full}
\end{equation}
In most practical situations there is symmetry, both on the inside of the
object that will be projected as on the outside. For example,
\begin{equation}
  A(p_1^{\,\mu_1} p_1^{\,\mu_2} p_1^{\,\mu_3} p_1^{\,\mu_4} p_2^{\,\mu_5} p_2^{\,\mu_6}) 
  p_3^{\,\mu_1} p_3^{\,\mu_2} p_4^{\,\mu_3} p_4^{\,\mu_4} p_4^{\,\mu_5}  p_4^{\,\mu_6}
  \label{eq:rank6ex}
\end{equation}
is symmetric in exchanges of $\mu_1,\ldots,\mu_4$ and $\mu_5,\mu_6$ inside $A$, and is symmetric in 
$\mu_1,\mu_2$ and $\mu_3,\ldots,\mu_6$ outside $A$.
The symmetry inside the object $A$ will enforce that coefficient $c_1$ and $c_2$ (and others) will 
actually be the same. The symmetry on the outside will cause terms to merge. In the end, we could 
have used the symmetrised variant of eq. \eqref{eq:rank6full} instead:
\begin{equation}
  c_1 \cdot (g^{\mu_1 \mu_2}g^{\mu_3 \mu_4}g^{\mu_5 \mu_6} + 2 g^{\mu_1 \mu_3}g^{\mu_2 \mu_4}g^{\mu_5 \mu_6}) 
  + c_3 (2 g^{\mu_1 \mu_2}g^{\mu_3 \mu_5}g^{\mu_4 \mu_6} + 10 g^{\mu_1 \mu_5}g^{\mu_2 \mu_6}g^{\mu_3 \mu_4} ) .
\label{eq:rank6sym}
\end{equation}
We see that only two coefficients have to be computed instead of 15 and that there are only 4 terms
in the output instead of 15. The challenge is to prevent these terms from being created in the 
first place by exploiting symmetry, instead of starting from eq. \eqref{eq:rank6full}.
We make use of the optimised \FORM{} command \texttt{dd\_}, which creates the tensor structure 
$T^{\mu_1,\ldots,\mu_r}$ without generating duplicates. 
If we evaluate \texttt{dd\_(p1,p1,p1,p1,p2,p2)} and strip the coefficient we get
\path{p1.p1^2*p2.p2+p1.p1*p1.p2^2}. 
These two terms represent the structure outside of $c_1$ and $c_3$ in \eqref{eq:rank6sym}. 
For each of these two terms, we solve for the coefficient.
Next, we recreate the metric structures that would give this specific contraction.

\vspace*{0.5mm}
A term generated by \texttt{dd\_} consists of two different factors: $(p \cdot p)^a$ and 
$(p_1^{} \cdot p_2^{})^a$. For $(p \cdot p)^a$, we collect all possible indices involved with $p$. 
For eq.~\eqref{eq:rank6ex}, this would be $\mu_1,\ldots,\mu_4$. Then we select all possible ways 
to get $2a$ elements from that list with \texttt{distrib\_}. Next, we use $dd\_$ on those indices. 
Thus, for $p_1^{} \cdot p_1^{}$ in the example we would get 
$g^{\mu_1 \mu_2}+g^{\mu_1 \mu_3}+g^{\mu_2 \mu_3}$. For cases such as $(p_1^{} \cdot p_2^{})^a$,
we select $a$ from the list of indices associated to $p_1^{}$ and $a$ from the list of $p_2^{}$. 
Then we permute over the list of $p_2^{}$.
Using this algorithm, one can generate all possible contractions from the result without 
generating duplicates.
To apply the outside symmetry, one can easily fill in the outside momenta associated to the 
indices instead of the indices themselves. 
\texttt{distrib\_} and \texttt{dd\_} will take the symmetry into account automatically.

\section{Five-loop results}
\label{sec:results}
\vspace*{-1mm}

\nin
In this section we briefly discuss the five-loop results for the beta function 
and the Higgs-boson decay to gluons.
With the same setup, we have also computed the Higgs decay to bottom quarks and 
the electromagnetic $R$-ratio at five loops \cite{Herzog:2017dtz}.  
Since these were re-computations, see ref.~\cite{Baikov:2005rw} and 
ref.~\cite{Baikov:2012er} and references therein, respectively, we will not 
address these results here.

\subsection{The beta function}
\label{sec:beta}
\vspace*{-1mm}

\nin
Using the method described above, we have computed the five-loop beta function
of QCD (obtained before in ref.~\cite{Baikov:2016tgj}) and its generalizations
to a general compact gauge group in the standard \MSb\ scheme, using the 
background field method in the Feynman gauge \cite{Herzog:2017ohr}. 
See refs.~\cite{Luthe:2017ttg,Chetyrkin:2017bjc} for more recent calculations 
with gauge dependence.
The analytic result in terms of rational numbers and the values 
$\zeta_{\,n\,}^{}$, $n = 3,4,5$ of the Riemann $\zeta$-function can be found in 
eq.~(3.5) of ref.~\cite{Herzog:2017ohr}.

\vspace*{1mm}
The \MSb\ result can be transformed to the MiniMOM (MM) scheme 
\cite{vonSmekal:2009ae,Gracey:2013sca}, which may be more convenient for 
extending analyses of the strong coupling constant into the non-perturbative
regime, using the {\sc Forcer} calculations of four-loop vertex functions in 
ref.~\cite{Ruijl:2017eht}. For the gauge-dependent general result see eq.~(B.4) 
of ref.~\cite{Ruijl:2017eht}.
It is interesting to note, e.g., in the context of 
refs.~\cite{JM-no-pi2,DV-no-pi2}$\,$%
\footnote{$\,$%
It appears that Euclidean physical quantities do not receive 
even-$n$ $\zeta_{\,n}$, i.e., $\pi^2$, contributions in renormalization schemes, 
such as MiniMOM or the scheme suggested in ref.\cite{Boito:2016pwf},
in which the beta function is free of such terms. 
Beyond the cases covered in refs.~\cite{JM-no-pi2,DV-no-pi2}, this has also 
been established to N$^4$LO \cite{KostjaPC} for the scheme-independent 
versions, see refs.~\cite{Chetyrkin:2004mf,Chetyrkin:2017bjc} of all 11 
propagator and vertex functions computed in ref.~\cite{Ruijl:2017eht}.
}, 
that the MiniMOM beta function in the Landau gauge, unlike the \MSb\ result, 
does not include $\z4$. The same was observed for the beta function
of QED in the MOM scheme in ref.~\cite{Baikov:2012zm}. For~further discussions 
of the issue of the $\zeta$-function values, see ref.~\cite{Baikov:2017ujl} 
and references therein.

\vspace*{0.5mm}
The numerical expansion of the \MSb\ beta function of QCD is very
benign to five loops with
\bea
\label{bQCD}
  \widetilde{\beta}(\als,\nf\!=\!3) & = &
       1
       + 0.565884 \,\* \als
       + 0.453014 \,\* \as(2)
       + 0.676967 \,\* \as(3)
       + 0.580928 \,\* \as(4)
       + \ldots
\; , \;\; \nn \\
 \widetilde{\beta}(\als,\nf\!=\!4) & = &
       1
       + 0.490197 \,\* \als
       + 0.308790 \,\* \as(2)
       + 0.485901 \,\* \as(3)
       + 0.280601 \,\* \as(4)
       + \ldots
\; , \;\; \nn \\
  \widetilde{\beta}(\als,\nf\!=\!5) & = &
       1
       + 0.401347 \,\* \als
       + 0.149427 \,\* \as(2)
       + 0.317223 \,\* \as(3)
       + 0.080921 \,\* \as(4)
       + \ldots
\eea
for $\widetilde{\beta} \equiv - \beta(a_{\rm s}) / (\ar(2) \beta_0)$
with $\beta_0 = 11 - 2/3\:\nf$ and $\ars = \als/(4 \pi)$.
The five-loop (N$^4$LO) contribution changes the beta function by less than 
1\% at $\als = 0.47$ for $\nf = 4$ and at $\als = 0.39$ for $\nf = 3$
quark flavours. 
The N$^{\,n \geq 2}$LO coefficients are larger in the Landau-gauge MiniMOM 
scheme,
\bea
\label{bMM}
  \widetilde{\beta}_\MM^{}(\als,\nf\!=\!3) & = &
    1
    + 0.565884 \,\* \als
    + 0.941986 \,\* \as(2)
    + 2.30450 \,\* \as(3)
    + 6.64749 \,\* \as(4)
    + \ldots
\; , \;\; \nn \\
 \widetilde{\beta}_\MM^{}(\als,\nf\!=\!4) & = &
    1
    + 0.490197 \,\* \als
    + 0.645215 \,\* \as(2)
    + 1.63846 \,\* \as(3)
    + 3.46687 \,\* \as(4)
    + \ldots
\; , \;\; \nn \\
 \widetilde{\beta}_\MM^{}(\als,\nf\!=\!5) & = &
    1
    + 0.401347 \,\* \als
    + 0.328852 \,\* \as(2)
    + 1.02689 \,\* \as(3)
    + 0.84177 \,\* \as(4)
    + \ldots
\eea
where $\als \equiv \as({\MM})$, and exhibit a definite growth with the order
that is absent in the \MSb\ case. 

\vspace*{0.5mm}
The different behaviour of the $\als$-expansion of the beta function 
of QCD in these two schemes is illustrated in the upper part of fig.~1.
At $\as({\MM}) = 0.25$, which corresponds to an \MSb\ value of $\als = 0.2$ 
for $\nf = 4$ at N$^4$LO -- for the conversion see eq.~(B.2) of 
ref.~\cite{Ruijl:2017eht} -- the individual N$^2$LO, N$^3$LO and N$^4$LO
contributions add 3.6\%, 2.3\% and 1.2\%, respectively, to the total
NLO result.
Unlike the \MSb\ case, where the expansion appears to converge up to rather 
large values of $\als$, the N$^3$LO contribution exceeds the N$^2$LO one 
for $\als \geq 0.4$, and the N$^4$LO effect that at N$^3$LO for 
$\als > 0.47$. 

\begin{figure}[t]
\vspace*{-1mm}
\centerline{\epsfig{file=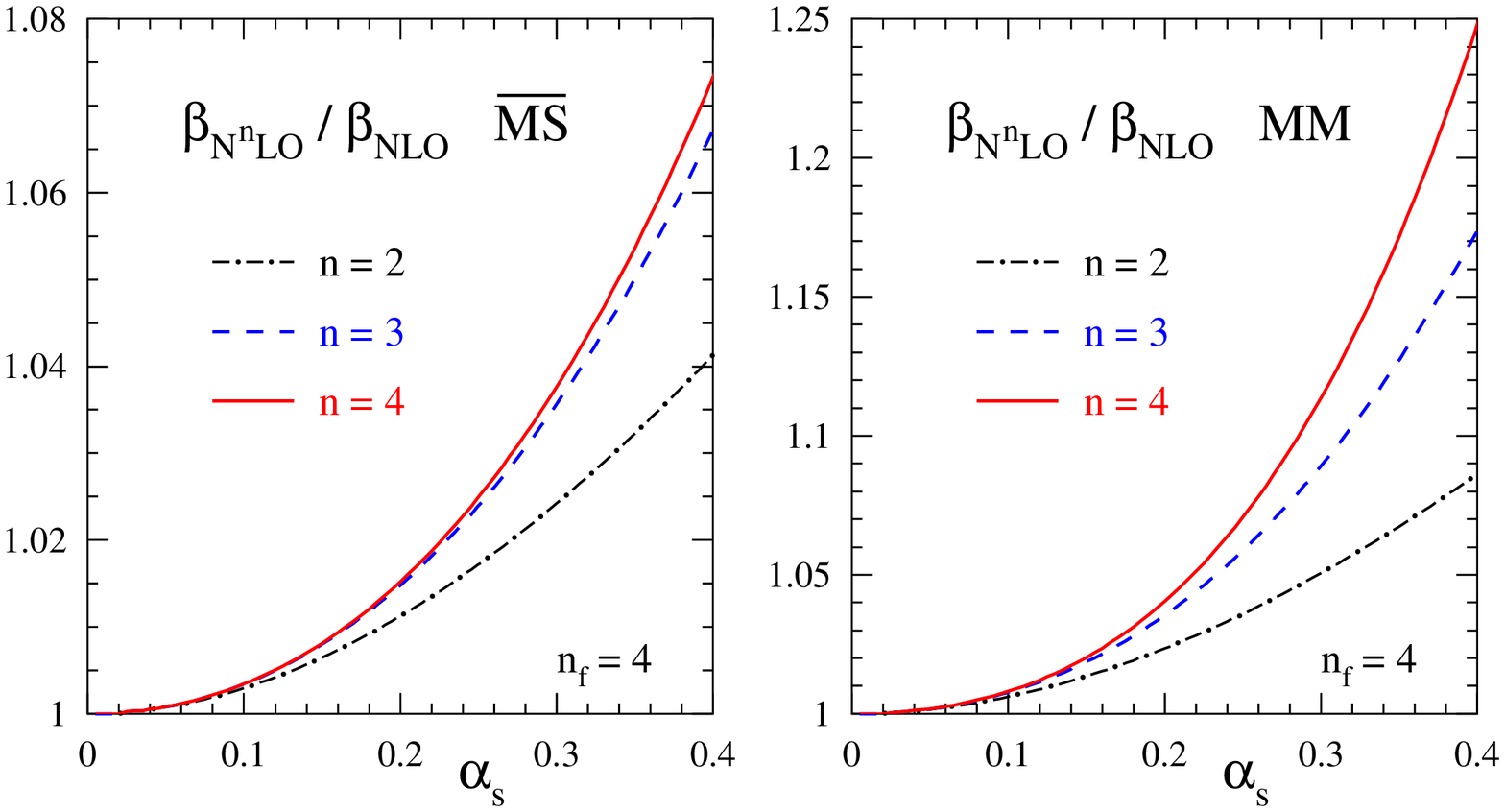,width=15.0cm,angle=0}} 
\centerline{\epsfig{file=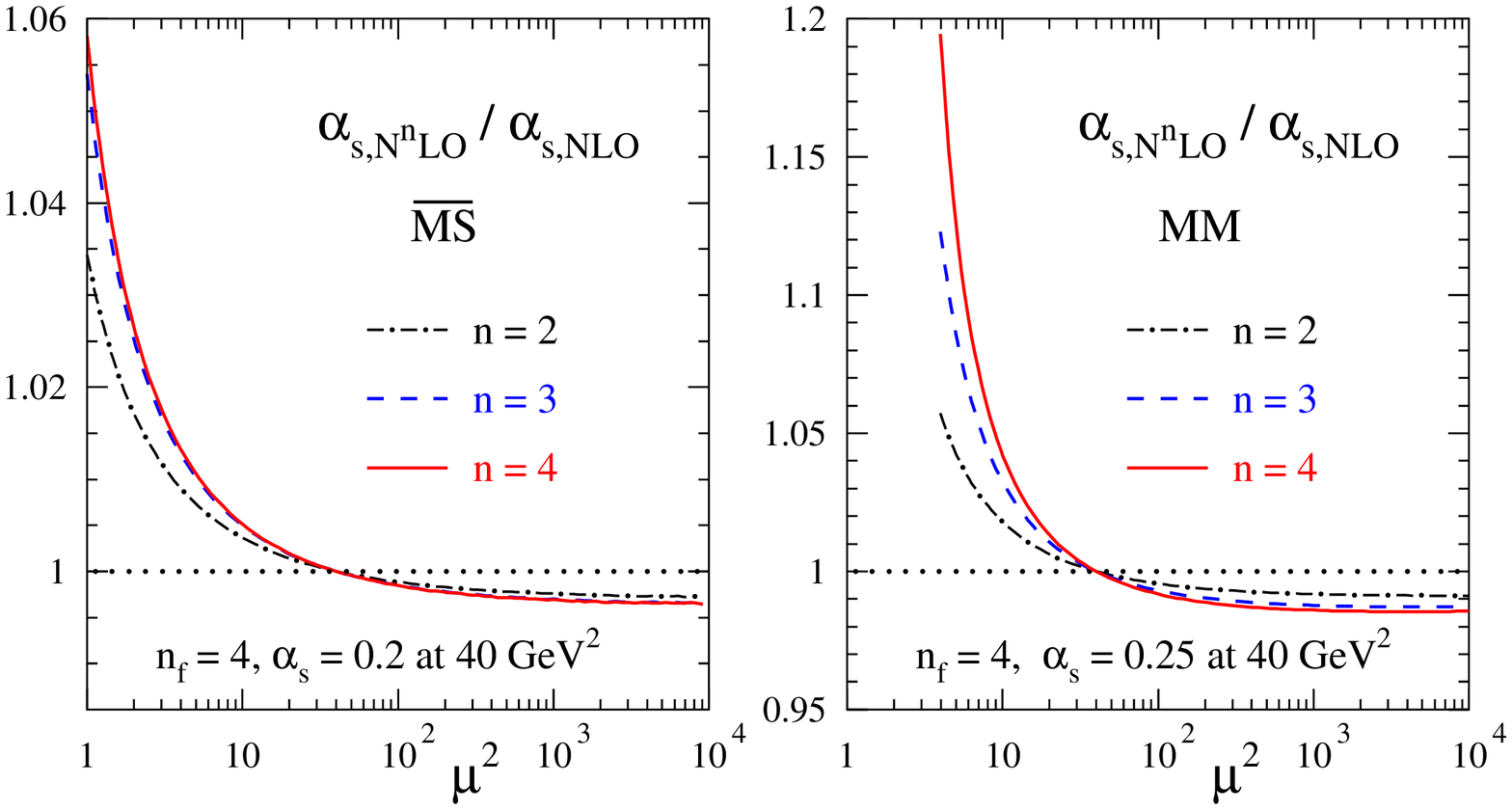,width=15.0cm,angle=0}}
\vspace{-2mm}
 \caption{ \label{fig1}
 Upper panels: the N$^2$LO, N$^3$LO and N$^4$LO approximation to the beta
 function of QCD with four flavours in the \MSb\ and MiniMOM (MM) schemes,
 normalized to their common NLO values.
 Lower panels: the resulting scale dependence of the respective coupling
 constants for order-independent reference values of 0.2 for \MSb\ and 0.25
 for MM at $\mu^2 = 40 \mbox{ GeV}^2$.
 }
\vspace*{-2mm}
\end{figure}

\vspace{0.5mm}
Hence, as illustrated in the lower part of fig.~1, also the running of $\als$
in the MiniMOM scheme
becomes unstable at~a scale of about $\mu \approx 2 \mbox{ GeV}$.  
For a comparison of the scale stability of the \mbox{$R$-ratio} in these two 
schemes at a moderate ($\als(Q^2) \simeq 0.2$ with $\nf = 4$ in \MSb) and a 
small ($\als(Q^2) \simeq 0.3$ and $\nf = 3$) scale, see figs.~5 and 6 of~%
ref.~\cite{Herzog:2017dtz}.

\subsection{Higgs decay to gluons}
\label{sec:higgs}

\nin
In the heavy-top limit, there is an effective coupling of the Higgs boson to
gluons. If the light-quark masses are neglected the Higgs decay to hadrons
can be expressed, using the optical theorem, as
\beq
  \Gamma_{H\to\, gg} \;=\; \frac{\sqrt{2}\, G_{\rm F}}{\MH}\: |C_1|^2 \,
  \mathrm{Im}\, \Pi^{\,GG}(-\MHs-i\delta) \;.
\eeq
The Wilson coefficient $C_1$, which includes the top-mass (scheme) dependence,
is known to N$^4$LO, see ref.~\cite{Chetyrkin:2016uhw} and reference therein.
Due to the analytical continuation from the spacelike case, only the pole 
part of Higgs-boson self-energy $\Pi^{\,GG}$ induced by the effective $Hgg$
coupling is required:
\beq
 \mathrm{Im}\, \Pi^{\,GG}(-\qs-i\delta) \;=\; \sin(L \pi \ep) \Pi^{\,GG}(\qs)
 \;=\; \Pi^{\,GG}(\qs) L \pi \ep + \ldots \; ,
\eeq
where $D = 4 - 2\:\!\ep$ and $L$ is the number of loops.
Consequently, $\mathrm{Im}\, \Pi^{\,GG}(-\qs-i\delta)$ and hence the N$^4$LO
decay rate can be computed using the setup described above. 
The computation is much more costly than that of the beta function and took 
almost two months on the machines available to us.

\begin{figure}[t]
\vspace*{-1mm}
 \centerline{\epsfig{file=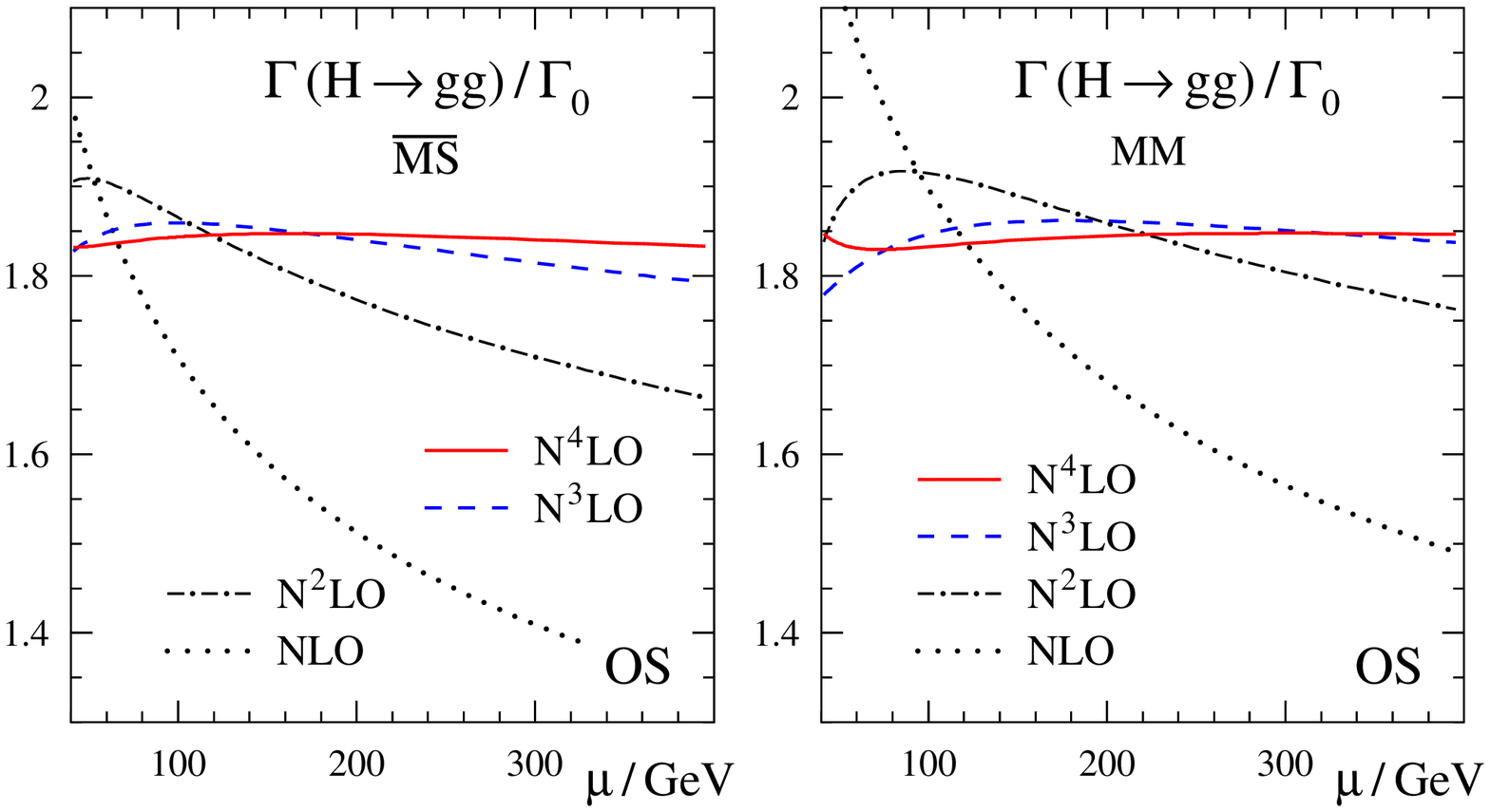,width=15.0cm,angle=0}}
\vspace{-4mm}
 \caption{ \label{fig2}
 The renormalization scale dependence of $\Gamma_{H \ra\,gg}$,
 normalized as described in the text, in the \MSb\ and MiniMOM schemes,
 for $\als(\MZ) = 0.118$ in \MSb, $\MH = 125 \mbox{ GeV}$ and an on-shell
 top-mass in 173 GeV.
 }
\vspace{1mm}
\end{figure}

\vspace*{0.5mm}
For an on-shell top mass of 173 GeV, the perturbative expansion of the 
decay width reads
\beq
\label{KOSnumMS}
 \Gamma_{H\to\, gg}^{\:\overline{\rm MS}} \;=\; \Gamma_0 \left(  
       1 + 5.703052\,\als    + 15.51204\,\as(2)
         + 12.6660 \,\as(3)  - 69.3287 \,\as(4) + \ldots \right)
\eeq
at $\nf = 5$ 
for the renormalization scale $\mu = \MH$ with ($G_F$ is the Fermi constant)
\beq
\label{GamHgg0}
  \Gamma_0 \;=\; G_F \MHt / (36 \pi^3 \sqrt{2}) \cdot (\als(\MHs))^2
\; .
\eeq
Note that the large ratio of the N$^4$LO and N$^3$LO coefficients is 
accidental, and not indicating a break-down of the perturbative 
expansion, see ref.~\cite{Herzog:2017dtz}.
The dependence of the numerical result on the scheme and value of
$m_{\,\rm top}$ is very weak. The corresponding MiniMOM result is given by
\beq
\label{KOSnumMM}
 \Gamma_{H\to\, gg}^{\:\rm MM} \;=\; \Gamma_0 \left(  
       1 + 4.345814\,\als    + 4.379443\,\as(2)
         - 21.5506 \,\as(3)  - 71.9231 \,\as(4) + \ldots \right)
\eeq
where, of course, $\als \equiv \as({\MM})$ with 
$\as({\MM})(\MZ) = 1.096\,\als(\MZ)$ for the standard \MSb\ value of 0.118.
This higher value is compensated by lower-order coefficients in 
eqs.~(\ref{KOSnumMM}) and that are smaller than their \MSb\ counterparts;
The N$^3$LO and N$^4$LO terms for $\nl=5$ are not smaller in MiniMOM, though.

\vspace*{0.5mm}
Taking into account also the renormalization scale dependence as shown in
fig.~\ref{fig2}, we arrive at 
\beq
\label{GHggRes}
  \Gamma_{\rm N^4LO} (H\ra gg) \;=\; \Gamma_0 \left(
  1.844 \:\pm\: 0.011_{\,\rm series} \:\pm\: 0.045_{\,\als(\MZ), 1\%}
  \right)
\eeq
with $\als(\MH) = 0.11264$, the \MSb\ value corresponding to 
$\als(\MZ) = 0.118$, in eq.~(\ref{GamHgg0}).
For $1/m_{\,\rm top}$ corrections and light-quark mass effects, see 
ref.~\cite{Davies:2017xsp} and references therein.
 
\section{Summary and outlook}
\label{sec:conclusion}
\vspace*{-1mm}

\nin
We have sketched how the $R^*$-operation can be used to compute the poles of Feynman diagrams.
Additionally, we have identified computational difficulties when performing five-loop calculations.
We provide four solutions that reduce the number of terms that are generated.

\vspace*{0.5mm}
Using these methods, we have computed the five-loop beta function for Yang-Mills theory with 
fermions~\cite{Herzog:2017ohr}, and the $R$-ratio and Higgs-boson decay widths $\Gamma_{\!H}$ to 
quarks and gluons~\cite{Herzog:2017dtz}. 
We~have briefly discussed the results for the beta function and for $\Gamma_{H\to\, gg}$ in the 
heavy-top limit. In the usual \MSb\ scheme, the perturbative running of the strong coupling 
constant $\als$ is now fully under control for all practical purposes. 
The uncertainty of $\Gamma_{H\to\, gg}$ due to the truncation of the perturbation
series is now much smaller than that due to present uncertainty of the value of $\als(M_Z)$;
measurements of a comparable accuracy are not possible at the LHC, but may be in reach of a future
$e^+e^-$ collider.

\vspace*{0.5mm}
It is possible to use our setup to extend recent calculations of splitting-function moments 
\cite{Ruijl:2016pkm,Moch:2017uml} to five loops for very low moments $N$.
The non-singlet splitting functions have already been computed in this manner at $N=2$ and $N=3$;
the results will be presented and discussed elsewhere.
The extension of these computations to $N \geq 4$ present a computational challenge well beyond
that posed by $\Gamma_{H\to\, gg}$, the by far hardest case presented here.

\subsection*{Acknowledgements}
\vspace*{-1mm}

\nin
We would like to thank K. Chetyrkin for useful discussions.
The research reported here has been supported by 
the {\it European Research Council}$\,$ (ERC) \mbox{Advanced} Grant 320651,$\,$%
{\it HEPGAME},
and the UK {\it Science \& Technology Facilities Council}$\,$ (STFC) grant
ST/L000431/1.


\providecommand{\url}[1]{#1}\providecommand{\href}[2]{#2}\begingroup\raggedright\endgroup
  
\end{document}